# Strong suppression of weak localization in graphene

S.V. Morozov[1,2], K.S. Novoselov[1], M.I. Katsnelson[3], F. Schedin[1], D. Jiang[1], & A.K. Geim[1]

[1]Department of Physics, University of Manchester, M13 9PL, Manchester, UK
[2]Institute for Microelectronics Technology, 142432 Chernogolovka, Russia
[3]Institute for Molecules and Materials, University of Nijmegen, 6525 ED Nijmegen, The Netherlands

*Low-field magnetoresistance is ubiquitous in low-dimensional metallic systems with high resistivity and well understood as arising due to quantum interference on self-intersecting diffusive trajectories. We have found that in graphene this weak-localization magnetoresistance is strongly suppressed and, in some cases, completely absent. The unexpected observation is attributed to mesoscopic corrugations of graphene sheets which can cause a dephasing effect similar to that of a random magnetic field.*

Graphene is a single layer of carbon atoms densely packed in a honeycomb lattice, or it can be seen as an individual atomic plane pulled out of bulk graphite. This material was found in its free state only recently, when individual graphene samples of a few microns in size were isolated by micromechanical cleavage of graphite [1]. The current intense interest in graphene is driven by both the unusual physics involved and a realistic promise of device applications. Two major features of graphene are largely responsible for the interest. First, despite being only one atom thick and unprotected from the immediate environment, graphene exhibits high crystal quality and ballistic transport at submicron distances [1,2]. Second, quasiparticles in graphene behave as massless Dirac fermions so that its electronic properties are governed by the physics of quantum electrodynamics rather than the standard physics of metals based on the (non-relativistic) Schrödinger equation (see [2,3] and references therein). Among relativistic-like phenomena observed in graphene so far, there are two new types of the integer quantum Hall effect and the presence of minimal metallic conductivity of about one conductivity quantum, $e^2/h$ [2-4]. The latter observation also means that there is no strong (Anderson) localization in graphene, and the material remains metallic even in the limit where concentrations of its charge carriers tend to zero.

In this paper, we report magnetoresistance (MR) measurements in graphene in the opposite, strongly metallic regime where quantum interference corrections to conductivity are widely expected to recover [5-8]. Indeed, despite the absence of strong localization, the interference on time-reversal quasiparticle trajectories seems unavoidable in the strongly metallic regime, and the recent theoretical analysis has predicted the standard magnitude for the quantum corrections (within a factor of 2), whereas their sign is generally expected to be positive (i.e. graphene should exhibit weak antilocalization) [5]. More recently [6,7], it has been argued that the presence of very-short (atomic) range scatterers can change this, leading to the possibility of both signs of weak localization corrections and even to their complete suppression, depending on disorder and temperature. In contrast, our experiments have shown only negative MR with a typical magnitude of two orders smaller than expected (at all temperatures). We have ruled out both a short phase-breaking length $L_\phi$ and magnetic impurities as possible mechanisms for the weak localization (WL) suppression. The unexpected behavior is most likely connected to mesoscopic corrugations (ripples) of graphene sheets, which were observed by atomic force microscopy (AFM) in many samples. We show that such distortions can indeed suppress quantum corrections because they lead to a fluctuating position of the Dirac point, which may be viewed as exposure of graphene to a random magnetic field [7,9]. The reason for always-negative MR remains to be understood.

The samples studied in this work were made from single-layer graphene flakes of several microns in size, which were placed on top of an oxidized silicon wafer (300 nm of $SiO_2$). A number of Au/Cr contact leads were attached to graphene sheets by using electron-beam lithography (inset in Fig. 1). To induce charge carriers in graphene we applied a gate voltage $V_g$ up to ±100V between graphene and the Si wafer, which resulted in carrier concentrations $n = \alpha V_g$ due to the electric field effect. The coefficient $\alpha \cong 7.2 \times 10^{10}$ cm$^{-2}$/V is determined by the geometry of the resulting capacitor and in agreement with the values of $n$ found experimentally from Hall effect measurements. For details of microfabrication and characterization of graphene devices, we refer to the earlier work [1,2,10].

Figure 1 shows one of our devices and changes in its resistivity $\rho$ with changing $V_g$. For a fixed gate voltage (i.e. fixed $n$), we measured changes in longitudinal resistivity $\rho_{xx}$ as a function of applied perpendicular field $B$. Examples of MR curves are plotted in Fig. 2a. The major anomaly on these curves is the fact that – away from the neutrality point $|n| \approx 0$ (see Fig. 2b) – they do not show any sign of positive or negative MR. This is striking because for metals with so high resistivity ($\approx$1kOhm per square), interference corrections should be significant and easily seen on the scale of Fig. 2. To emphasize this fact, we show the MR behavior normally expected [11,12] for a metallic film of the same resistivity under similar conditions. As further evidence for the anomalous behavior of graphene, Fig. 2c plots MR observed in multilayer graphitic films (about 10 atomic layers in thickness), which exhibit the WL behavior well described by the standard theory [11,12]. It is clear that for some reasons WL in single-layer graphene is strongly suppressed. This report concentrates on the strongly metallic regime ($n > 10^{12}$cm$^{-2}$) that has been the focus of recent theory [5-8] but, for completeness, Fig. 2 also shows the magnetoresistance of graphene in the region of $|n| \approx 0$. No hint of WL magnetoresistance was observed in this regime either. Instead, we usually saw a large positive MR (Fig. 2b), which behaves as $B^2$ with characteristic fields $B > 1$T. This low-$n$ MR was essentially temperature-independent, indicating its non-interference origin, and can be explained by standard classical effects due to the presence of two types of charge carriers [12]. The positive MR gradually faded away with increasing $n$.

The behavior shown in Fig. 2a was rather common (exhibited by >80% of our samples) and observed at temperatures $T$ from liquid nitrogen down to 0.3K. However, in some cases, we did see a small negative MR peak at zero $B$, which had the same shape as expected for WL but a much smaller height. By studying this remnant magnetoresistance in detail, we were able to narrow the range of possible explanations. Figure 3 shows MR for one of a few samples where the remnant peak was relatively large. By measuring its $T$-dependence, we found that although the peak's height $\Delta \rho$ was 10 times smaller than expected, it varied as $\ln(T)$, which is distinctive for quantum-interference corrections in two dimensions. Also, by fitting the shape of the MR peak using the standard formulas [11,12], we determined the phase-breaking length $L_\phi$ and its temperature dependence (see insets in Fig. 3). $L_\phi$ varied approximately as $1/\sqrt{T}$ and reached $\approx 1 \mu$m at 4K. The general theory of phase randomization processes in metallic systems allows [11] an estimate for the phase-breaking time $\tau_\phi$ as $\hbar/\tau_\phi \approx T/k_F l$, which leads to $L_\phi \approx l(E_F/2T)^{1/2}$ where $E_F$ and $k_F$ are the Fermi energy and wavevector of Dirac fermions, respectively. Both the $T$-dependence and absolute values of $L_\phi$ are in good agreement with the theory. These observations rule out electron heating or any other uncontrollable inelastic mechanisms as the reason for the WL suppression in our experiments.

The magnetoresistance traces in Figs 2 and 3 also show pronounced fluctuations, which were reproducible and identified as universal conductance fluctuations (UCF). Unlike WL, the mesoscopic fluctuations did not exhibit any anomaly in the metallic regime $k_F l >> 1$: their correlation field yielded the same values of $L_\phi$ as found from the WL analysis, and the UCF amplitude was in agreement with theory (i.e. $\approx e^2/h$, after taking into account the averaging over different phase coherent regions). Furthermore, the behavior of UCF indicated no spin-flip scattering in graphene. Indeed, the interaction of electrons with localized spins is known to suppress UCF, whose amplitude then becomes a strong, exponential function of $B$ [13], whereas in our experiments the fluctuations were essentially independent of $B$ for all $T$. Moreover, the conventional WL magnetoresistance observed in multilayer graphitic films (prepared under the same conditions as graphene) makes magnetic impurities highly improbable as the origin for the suppression of WL in graphene.

The existing theories [5-8] may perhaps explain the suppressed WL by interplay between localization and antilocalization caused by different types of defects, which fortuitously cancel each other. However, not a single one of our samples (several dozens were studied) has shown any sign of positive WL magnetoresistance at any temperature, which makes such a model implausible. To explain our results, we first note that graphene samples were often found to have an undulating surface, as shown in the inset of Fig. 1. The height $Z$ of these ripples could be up to several Å and they were typically a few tens nm in lateral size $d$. Smaller or sharper ripples are also possible but their detection is beyond the AFM resolution (we used Nanoprobe III). We believe that the observed ripples appear during micromechanical cleavage [1,2]. In this process, released graphene flakes are unlikely to be absolutely flat and cannot simultaneously attach to a Si substrate over their entire surface, which should lead to wrinkling.

The mesoscopic ripples can cause local elastic distortions, which effectively result in a random gauge field $A$ (leading to the replacement $i\hbar\nabla \rightarrow i\hbar\nabla + A$), following the mechanism first proposed by Iordanskii and Koshelev for the case of dislocations in multivalley conductors [9]. This gauge field breaks down the time-reversal symmetry in the vicinity of the Dirac points [7,9], which leads to suppression of the normal WL behavior. Applying the earlier analysis to our particular case, corrugations in graphene can be described by the tensor $\delta\bar{u}_{ij} = \frac{1}{2}\frac{\partial Z}{\partial x_i}\frac{\partial Z}{\partial x_j}$ and lead to changes in the nearest-neighbor hopping integral $\gamma_0(x_i, x_j) = \gamma_0(0) + \left(\frac{\partial \gamma_0}{\partial \bar{u}_{ij}}\right)\delta\bar{u}_{ij}$. Because the integral $\gamma_0$ becomes a function of in-plane coordinates $x_i$ and $x_j$, this results in shifts of Dirac points $K$ and $K'$, which in turn is equivalent to applying a field with amplitude $A \approx \hbar\gamma_0|\nabla Z|^2/v_F$, where $v_F \approx 10^6$ m/s is the Fermi velocity of Dirac fermions. The distortion gradient $\nabla Z$ can be estimated as $\approx Z/d$. The above expression yields that our graphene films should effectively behave as if they were exposed to a random local field $b$ of $\approx 0.1$ to 1T. A typical non-compensated flux $\Phi$ induced by the random field inside a phase coherent trajectory of size $L_\phi$ is given by $\Phi \approx b(L_\phi d)$ and exceeds one flux quantum under most conditions in our experiments, in agreement with complementary estimates in ref. [7]. It is important to mention that field $A$ has opposite signs for $K$ and $K'$ valleys so that there is no violation of the time-reversal symmetry for wrinkled graphene as a whole (for example, ripples cannot cause the Hall effect because contributions from two valleys cancel each other). However, in the absence of Umklapp processes, the electron subsystems near $K$ and $K'$ points are effectively independent, and the gauge field induced by ripples destroys quantum interference in the same way as magnetic field [7,9].

The discussed ripples allow one to understand the entire experimental picture self-consistently. Indeed, the inferred values of $b$ are sufficient to explain the complete suppression of WL in strongly rippled graphene at all temperatures. On the other hand, if a sample has only a partial coverage with such ripples, this should lead to a reduced height of its WL peak (we did see some correlation between the amount of ripples and the height of the WL peak). At the same time, a random magnetic field should not affect mesoscopic fluctuations (UCF), in agreement with the experiment. Ripples are also expected to become smaller in thicker and more rigid graphitic films, in agreement with our AFM observations. This is consistent with the fact that no anomalies were found in the WL behavior of our multilayer devices.

To summarize, both universal conductance fluctuations and weak localization are absent in graphene at low concentrations of Dirac fermions ($k_F l \approx 1$) but UCF fully recover in the metallic regime $k_F l >> 1$ whereas WL is found to remain strongly suppressed. The observed remnants of WL magnetoresistance were always negative, which appears to disagree with the existing theoretical models. As for the WL amplitude, its observed suppression is also unexpected, and we attribute it to the presence of mesoscopic ripples. Such ripples should certainly be taken into account in further studies of graphene and in trying to improve its mobility. To this end, WL magnetoresistance can be used as an indication of graphene's quality. On the other hand, rippled graphene can be used to address certain cosmological analogies [14] and offers an opportunity to study the physics associated with transport in random magnetic fields, a problem that was intensively discussed theoretically during the last decade but difficult to access experimentally for other systems [15].

Acknowledgements: We are grateful to Boris Altshuler, Carlo Beenakker, Antonio Castro Neto, Vladimir Falko, Paco Guinea, Dmitri Khveshchenko, Leonid Levitov, Allan Macdonald and Klaus Ziegler for illuminating discussions and comments. This work was supported by EPSRC (UK).

*Note added in proof.* – Most recently, to improve the quality of our graphene samples, we attempted to eliminate the mesoscopic ripples discussed in the paper. To this end, we have changed our microfabrication procedures [1] by depositing flakes on the freshly-cleaned $SiO_2$ surface (within 1 hour). This technological change resulted in samples with generally higher mobility (up to 15,000 cm$^2$/Vs) with no ripples visible in AFM. Moreover, such structures exhibited the full, unsuppressed WL peak. The experimental curves look very similar to the one shown in Fig. 3 but no additional fitting parameter is required to explain negative MR peak's amplitude. This proves that the WL amplitude (but not its sign) is sensitive to fabrication procedures and further supports the inferred importance of ripples.

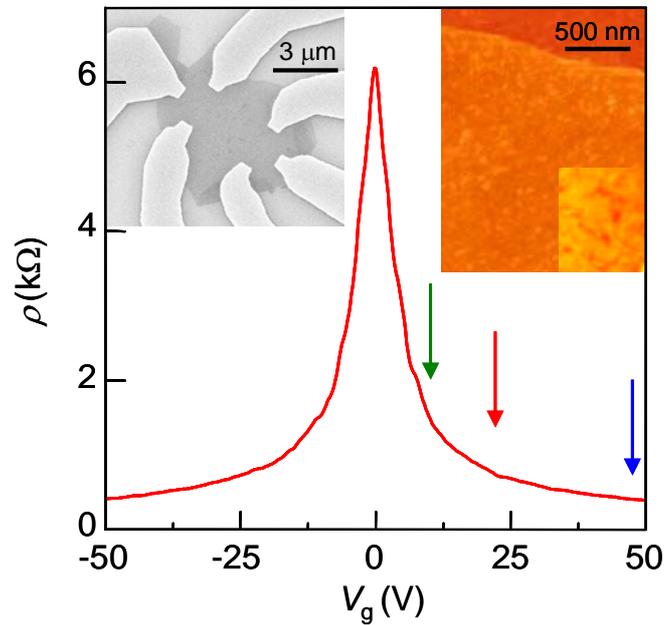

Figure 1. (color online). (Right inset) – High-resolution AFM image of a graphene flake before microfabrication. The top (darker) part of the image, where no ripples are visible, is an oxidized Si wafer (the step height is ≈6Å). The smaller inset shows a 3 times magnified AFM image with contrast enhanced in order to see the ripples clearer. (Left inset) – Scanning electron micrograph of one of our devices. Some larger ripples can also be seen on electron microscopy images. (Main panel) – Changes in resistivity of graphene with changing gate voltage. The arrows indicate gate voltages that correspond to magnetoresistance traces in Fig. 2a.

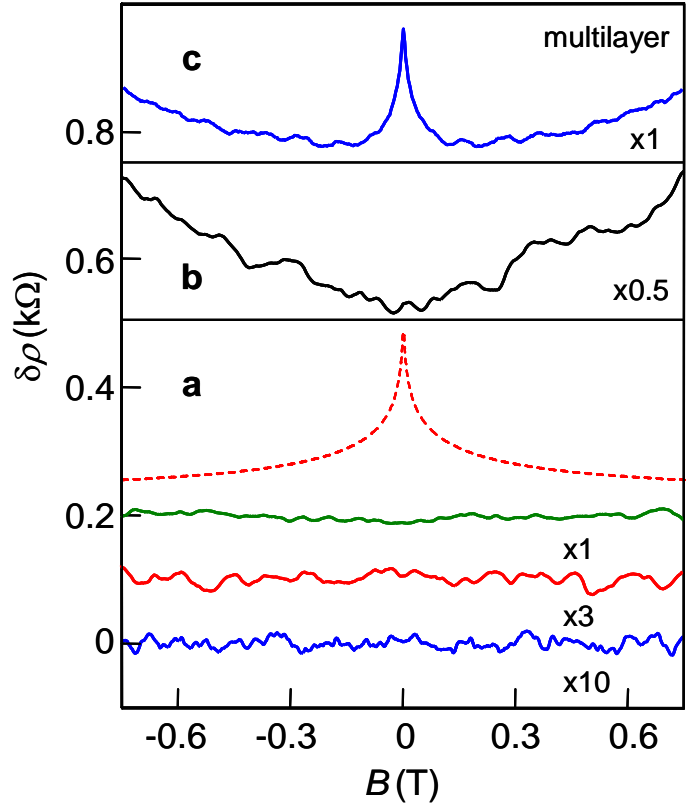

Figure 2. (color online). (a) – Many graphene devices exhibited no sign of weak localization or antilocalization. Solid curves correspond to gate voltages shown by arrows in Fig. 1 ($\approx$10, 20 and 50V from top to bottom curve, respectively). The curves are shifted for clarity ($\rho \approx$1.5, 0.8 and 0.4 k$\Omega$ from top to bottom). The lowest curve corresponds to $k_F l \approx$50. Notice magnification factors for the $\delta\rho$-scale against each of the curves. These factors were chosen so that the expected WL peak for all the curves would be of approximately the same size as the peak shown by the dashed curve calculated using the standard WL theory [11,12]. (b) – Magnetoresistance behavior at zero $V_g$ where $\rho$ reaches its maximum $\approx$6k$\Omega$. For so high resistivity (i.e. $\approx h/e^2$ per each type of carriers), a metal-insulator transition is generally expected but it does not occur in the case of graphene [2]. Both absolute value of $\rho$ and its magnetoresistance $\delta\rho(B)$ are practically temperature-independent below 100K. (c) – Multilayer films [10] exhibited the standard weak localization behavior. Shown is a device with $\rho \approx$1.2k$\Omega$ and mobility $\approx$10,000 cm$^2$/Vs (no gate voltage applied). A clear WL peak is seen at zero $B$. In higher fields, multilayer devices exhibit a large linear ($\propto B$) magnetoresistance. All the curves shown in Fig. 2 were measured at 4K.

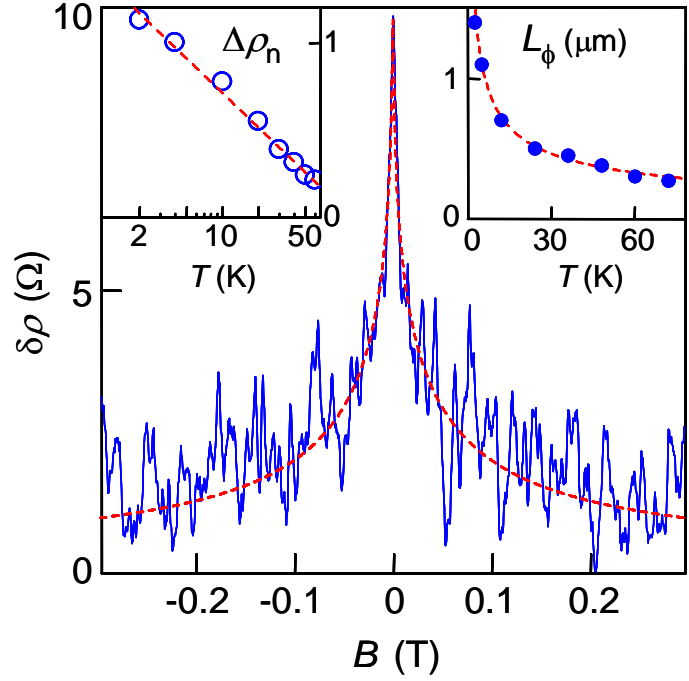

Figure 3. (color online). A graphene device exhibiting some remnants of weak localization. The main panel shows its low-field magnetoresistance (solid curve; $n \approx 3 \times 10^{12} cm^{-2}$; $l \approx 80 nm$; $T = 4K$). The dashed curve is the standard theory [11,12] but scaled along the y-axis by a factor of 0.11 in order to fit the experimental curve. Because $L_\phi$ can also be found from the correlation field of UCF, the absolute amplitude of the WL peak is the only fitting parameter. (Left inset) – Height $\Delta\rho$ of the MR peak as a function of $T$. Symbols are experimental data (normalized to $\Delta\rho$ at 4K); $\ln T$ dependence is shown by the dashed line. (Right inset) – phase-breaking length $L_\phi$ (symbols) is well described by $1/\sqrt{T}$-dependence (dashed curve).